\documentclass[letterpaper, 10 pt, conference]{ieeeconf}  

\usepackage{graphicx}
\IEEEoverridecommandlockouts                              
\begin{document}
\title{\LARGE \bf Synchronizing Geospatial Information for Personalized Health Monitoring}


\author{Nitish Nag$^{1,3}$, Vaibhav Pandey$^{1}$, Likhita Navali$^{2}$, Prateek Mohan$^{2}$, Ramesh Jain$^{1}$
\thanks{*This work was not supported by any organization}
\thanks{$^{1}$ Donald Bren School of Information and Computer Science,
        University of California, Irvine, United States of America
        {\tt\small vaibhap1 at uci.edu}}%
\thanks{$^{2}$ Department of Computer Science PES University,
        Bangalore, India
        {\tt\small email at likhitanavali at gmail.com and prateekmohan1997 at gmail.com}}%
 \thanks{$^{3}$ Medical Scientist Training Program, School of Medicine, University of California, Irvine, United States of America,
        {\tt\small nagn at uci.edu}}%
}

\maketitle
\begin{abstract}
The health effects of air pollution have been subject to intense study in recent decades. Exposure to pollutants such as airborne particulate matter and ozone has been associated with increases in morbidity and mortality, especially with regards to respiratory and cardiovascular diseases. Unfortunately, individuals do not have readily accessible methods by which to track their exposure to pollution. This paper proposes how pollution parameters like CO, NO2, O3, PM2.5, PM10 and SO2 can be monitored for respiratory and cardiovascular personalized health during outdoor exercise events. Using location tracked activities, we synchronize them to public data sets of pollution sensors. For improved accuracy in estimation, we use heart rate data to understand breathing volume mapped with the local air quality sensors via constant GPS tracking.
\end{abstract}

\section{INTRODUCTION}
Air pollution is a major environmental risk to health. By reducing air pollution levels, countries can reduce the burden of disease from stroke, heart disease, lung cancer, and both chronic and acute respiratory diseases, including asthma \cite{c3}.The lower the levels of air pollution, the better the cardiovascular and respiratory health of the population will be, both long- and short-term. Air pollutants in the local environment come into contact with individuals most during heavy outdoor cardiovascular exercise, such as cycling or running. Hence we target this particular situation in this paper. We consider the breathing rate as a major factor in the estimation of intake of pollutants during an exercise, because depending on the total tidal volume and also the breathing rate of an individual, the amount of pollutant intake changes. For example, a person having a normal breathing rate in a densely polluted area might be inhaling the same amount of pollutant as a person in a moderately polluted area with a very high breathing rate.

\section{RELATED WORK}
Air pollution is a major cause of death and diseases globally. An estimated 4.2 million premature deaths globally are linked to ambient air pollution, mainly from heart disease, stroke, chronic obstructive pulmonary disease, lung cancer, and acute respiratory infections in children \cite{c10}. Hence determining the after affects of air pollution on human health would help people to take precautionary measures. Pollutants with the strongest evidence for public health concern, include particulate matter (PM), ozone (O3), nitrogen dioxide (NO2) and sulfur dioxide (SO2) \cite{c10}.

Breathing rate and tidal volume are estimated from heart rate during exercise. The breathing rate and tidal volume vary in response to metabolic demand and increase in physical activity \cite{c11}.First, the tidal volume of the individual is calculated which is the lung volume representing the normal volume of air displaced between normal inhalation and exhalation. Tidal volume is calculated from the ideal body weight which requires the height, sex and the age of the individual. From the ideal volume the tidal volume is calculated \cite{c15}. An increase in ventilation can be affected by an increase in both the depth and frequency of breathing. Thus due to  the respiratory frequency increasing, and the tidal volume being at it's peak the total intake of the pollutants is calculated \cite{c16}.

The cigarette equivalent is derived from both the pollution intake and the tidal volume, where depending on the percentage of each of the pollutant intake, the total summation gives the result. Here the cigarette equivalent is used to  eliminate the confusion, by converting real-time air quality data from pollution levels into the equivalent number of cigarettes smoked over time. We use this for semantic understanding by the user. Since inhaling cigarette smoke has been shown to produce acute changes in the lung including alterations in resistance to airflow, cough, and irritation of the airway, the early stage of smoking might affect the respiratory function, converting the air quality to this standard gives a better insight of how the pollution affects any individual \cite{c13} \cite{c14}.

\section{Methodology}
 
\subsection{Data collection}

To correlate the pollution and health we collected publicly available data of air pollution data streams from United States government sensors. We obtained user permission and data from location based exercise application Strava \cite{c6}.

\subsubsection{Air pollution}

The level of pollutants like CO, NO2, SO2, PM2.5, PM10 and O3 were collected from Openaq \cite{c9}. It aggregates physical air quality data from public data sources provided by the government, research-grade and other sources. Openaq provides data in json format with location and time stamp. This is used to tag corresponding latitude, longitude and time with respect to each activity. Each of the pollutants is collected in  µg/m3 unit. Due to unavailability of the sensors to collect data in some places the exact pollutant level is not known hence the pollutant values of the nearest sensor is taken into consideration. There is significant potential for future research to improve this accuracy level.

\subsubsection{Health Parameters}
Strava tracks activities like cycling, running, swimming etc using GPS based wearable devices, which can include health parameters like heart rate and power. Users can record and upload physical activities and the software provides statistics such as distance, pace, heart rate, and elevation, and comparisons with activities of other users.

\subsubsection{Breathing Rate and tidal volume}

Breathing rate is derived from heart rate or power generated during exercise \cite{c5}. Tidal volume is the volume of air inhaled during a normal breath. It mainly depends on gender, weight and height of an individual. Estimation of the total volume of air intake would simply be the product of breathing rate and tidal volume. This is the feature used to calculate how much of each pollutant is inhaled during an exercise.

\subsection{Data Analysis}

The data analysis includes 
\begin{itemize}

\item Finding pollutant level using the following factors
\begin{itemize}
    \item latitude
    \item longitude
    \item time
    
\end{itemize} This allows us  to have an accurate measure of the constantly changing pollutants that are present and affect the user in the course of the exercise at different locations. This also allows for a more accurate reading as the pollution levels can be very different depending on the location and time.
\item Finding correlation between heart rate, power output (if available for the user activity) and the breathing rate.
\begin{itemize}
    \item The heart rate is considered as it allows us to have the best conversion to the breathing rate of any individual. During an exercise the  heart rate and breathing rate have a purpose in common: getting Oxygen. Due to higher demand of oxygen during an exercise the supply for the oxygen is increased by the heart pumping blood faster, hence having an increase in the heart rate. This shows that during a course of an exercise the heart rate can give a very mirrored view of the breathing rate.
    \item With heart rate being recorded the addition of power levels can amplify the correlations between higher heart rates resulting in higher breathing levels. As the breathing rate is increased for a higher intake of oxygen which increase the heart rate, we notice that the power levels are also increased. This follows the principle that the more oxygen intake is being used to spend as energy which results in higher power level. Also due to power level and heart rate being tightly linked, in case of the absence of heart rate, power level can solely be used to provide a better insight on the individuals breathing rate.
\end{itemize}
\item Finding the Cigarette Equivalence.
\begin{itemize}
    \item Calculating the total intake of each of the pollutants.
    \item Converting each of the different pollutants into a ratio that is equivalent to a cigarette. 
    \item summation of all the ratios from  the different pollutants which gives the total cigarette equivalent for any activity.
\end{itemize}
\end{itemize}
\begin{figure}[h!]
  \caption{Flowchart of the two main components}
  \centering
  \includegraphics[width=0.45\textwidth]{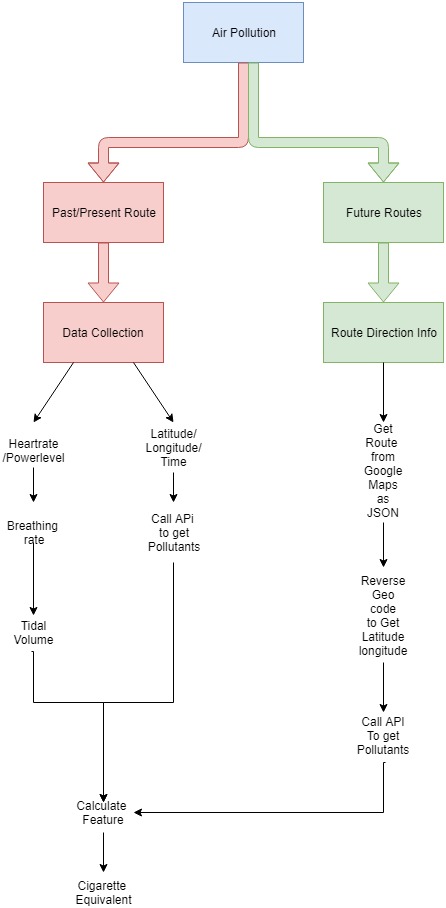}
Figure 1 is the model representation. It shows the flow of data from the user's Strava data, air-pollution data and Google maps API. It also shows how the feature is calculated and used to get the cigarette equivalent.
\end{figure}

\section{Maths}

\subsection{Feature Calculation}

Breathing rate is calculated from heart rate, if heart rate is not present then we use power as an alternative.

Tidal volume for men is calculated using the following formula:
\newline 

\textbf{\textit{volume= ((50 + 2.3*(height-60))*12)/1000 liters}}
\newline 

Tidal volume for women is calculated using the following formula:
\newline

\textbf{\textit{volume= ((45 + 2.3*(height-60))*12)/1000 liters}}
\newline

Feature is calculated by multiplying the breathing rate with tidal volume. This is the total volume intake.

\subsection{Calculation of pollutant intake}

Depending on the latitude, longitude and time the content of each pollutant is collected. Value of each pollutant is multiplied with the feature to get the pollutant intake.

\subsection{Calculation of pollutant intake for future rides}

The Google API returns json object from which road names are extracted and this is converted to the latitude, longitude by reverse
Geo-coding. This is used as a parameter to call upon our API which returns the level of air pollutants like CO, NO2, SO2, PM2.5, PM10 and O3.
Depending on the mode of transport heart rate will vary. Heart rate is directly proportional to the breathing rate, hence changing the pollutant intake drastically.

\begin{figure}[h!]
  \caption{Heart rate (blue) and breathing rate (orange) variation over time. }
  \centering
  \includegraphics[width=0.45\textwidth]{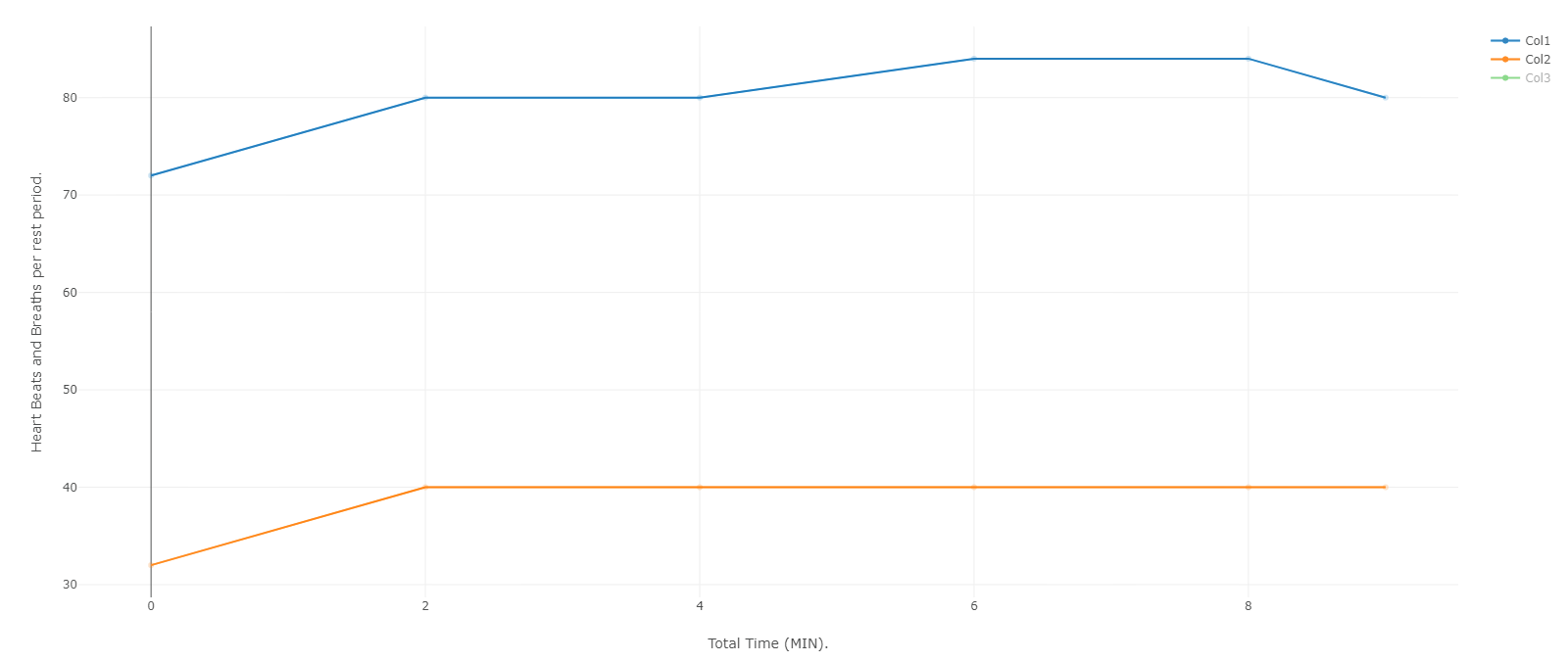}
\end{figure}

\section{Visualization}

\begin{figure}[h!]
  \caption{PM2.5 level for an activity}
  \centering
  \includegraphics[width=0.4\textwidth]{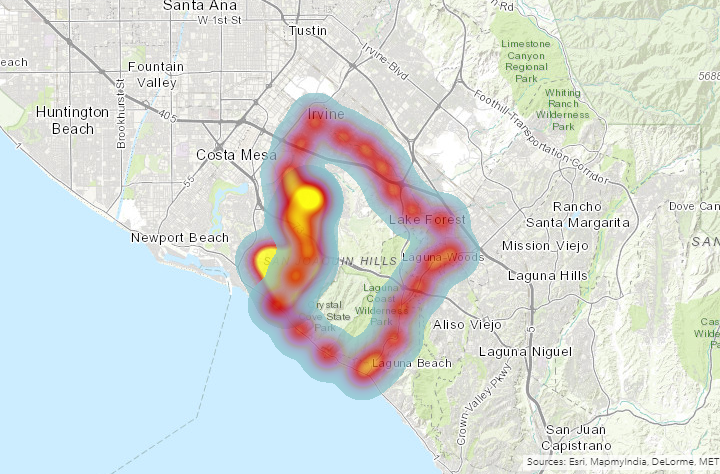}
\end{figure}

\begin{figure}[h!]
  \caption{CO level for an activity}
  \centering
  \includegraphics[width=0.4\textwidth]{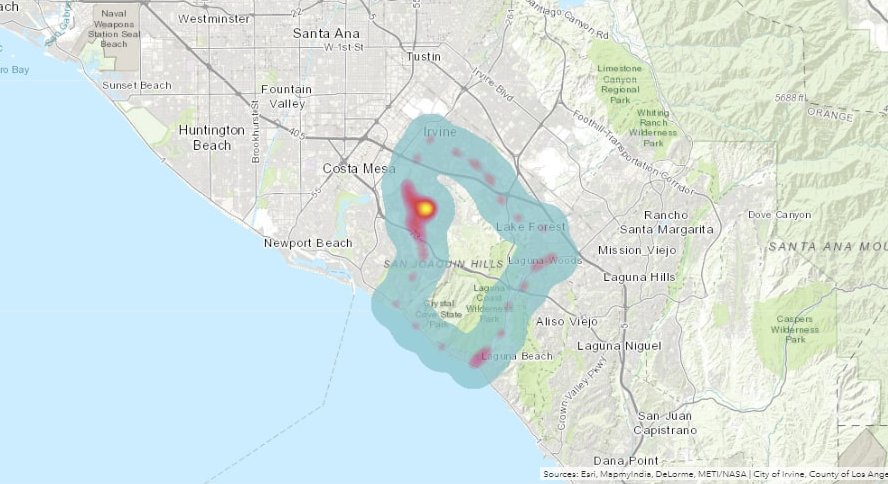}
\end{figure}

Comparing Figure 3 and Figure 4 we derive that the path taken by the rider has more of PM2.5 level compared to CO. The carbon monoxide level is 0.09 μg/m3 and that of PM2.5 is 6.4 μg/m3. These would allow the user to understand where their health is being affected, and potentially change future route choices.

\begin{figure}[h!]
  \caption{}
  \centering
  \includegraphics[width=0.4\textwidth]{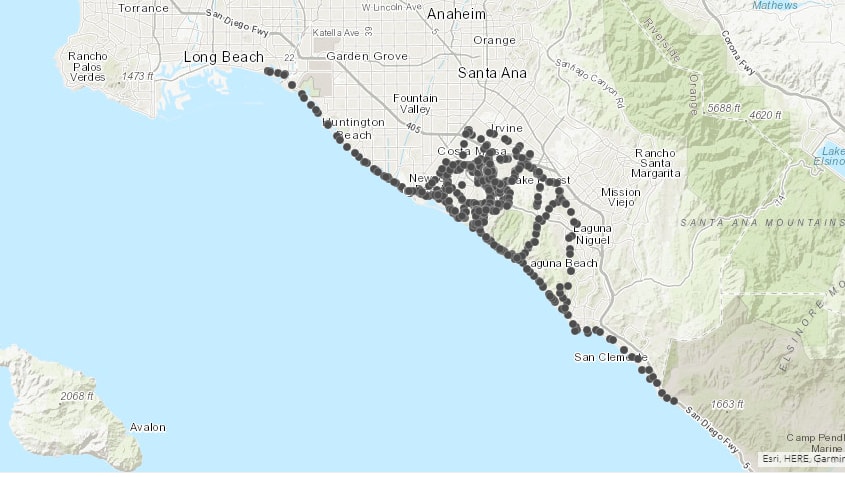}
\end{figure}

\begin{figure}[h!]
  \caption{.}
  \centering
  \includegraphics[width=0.4\textwidth]{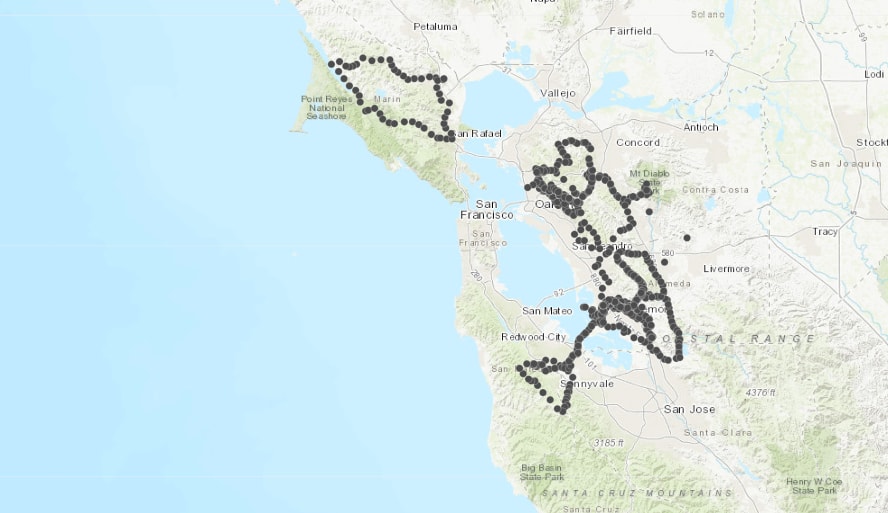}
\end{figure}

Figure 5 and Figure 6 indicate the location where the activities were Geo-coded. The data of the activities is formatted and plotted on ArcGIS software.

\begin{figure}[h!]
  \caption{O3 level for an activity}
  \centering
  \includegraphics[width=0.4\textwidth]{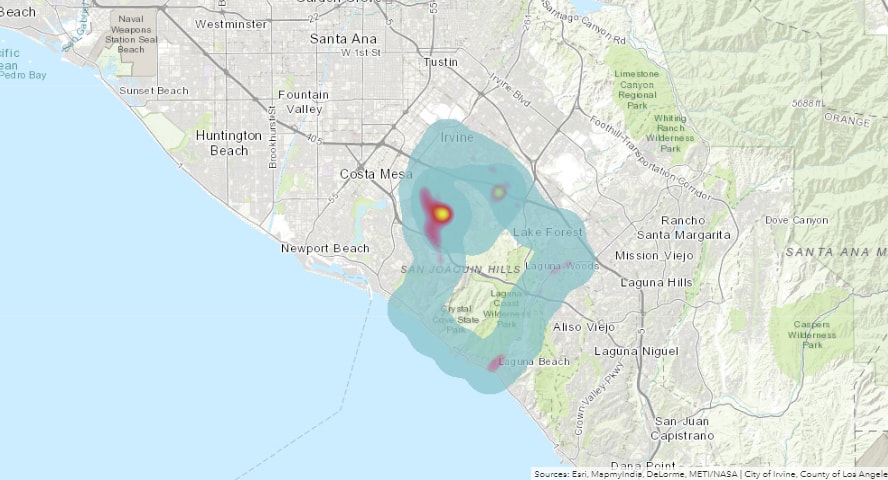}
\end{figure}

\begin{figure}[h!]
  \caption{PM10 level for an activity}
  \centering
  \includegraphics[width=0.4\textwidth]{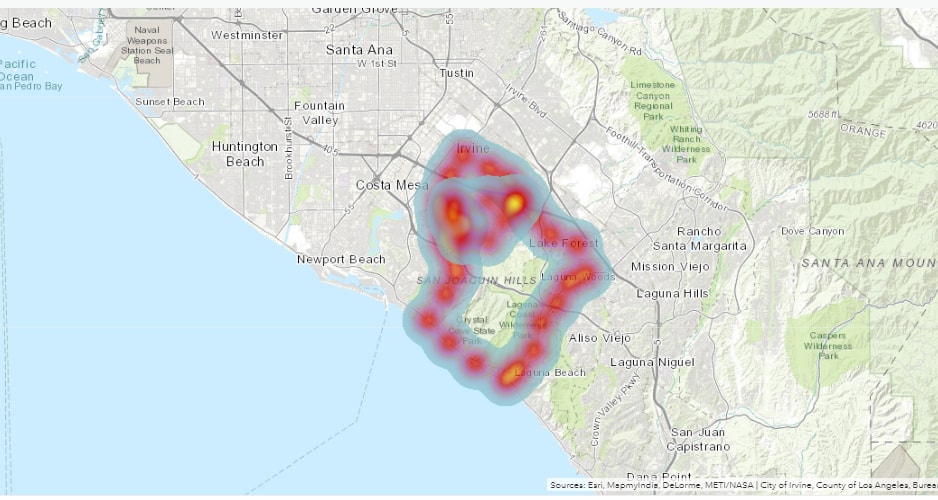}
\end{figure}

Figure 7 indicates the O3 content and Figure 8 indicates the PM10 content. It is clear that PM10 level is more than O3. The O3 level is 0.02 μg/m3 and that of PM10 is 6.0 μg/m3.

\section{Results}

Depending on the level of pollutant intake we group them into several concentration levels having different symptoms for each level as seen on the color heat map in Figures 3, 4, 7, and 8. Carbon-monoxide is very dangerous if inhaled more than the lethal level. 1ppm of carbon-monoxide is equivalent to one cigarette \cite{c4}. Similarly 10 μg/m3 of PM2.5 and NO2 and 1 μg/m3 of black carbon corresponds to an average of 5.5, 2.5, 4 passively smoked cigarettes per day respectively \cite{c1} \cite{c2}. 80 activities of an individual user were processed and it is observed that was the equivalent of smoking 3 cigarettes. These activities are bulk plotted in Figures 5 and 6.

\section{CONCLUSIONS}

This paper proposes mechanisms for personal health monitoring by using location data, health parameters, and public sensor data streams. This is only one component of the larger exposome that an individual would have computed with modern systems for health state estimation \cite{c19} \cite{c17}. This could be used to understand the causal reasons behind changes in respiratory or cardiovascular health by the user, a medical professional, or a computational system such as personal health navigators \cite{c18}. 

We hope to spark more advanced research in using geospatial data for personal health monitoring. Using this information may help in predicting the pollutant intake for future rides if used with real-time environmental sensor data streams. For users who are looking to understand how they can predict pollution effects on their next activity, future work can provide an estimation of pollution exposure given a location. A simple method for this could be to have the user input the origin and destination point along with the mode of transportation, primarily being walking, running, cycling, or driving. A call is placed to a routing API with parameters being the origin and destination, the API returns the directions. This can then be reverse geo-coded providing a list of latitudes and longitudes for the future route. From here we can give an estimated pollution exposure for that activity at the given time of query.

\addtolength{\textheight}{-12cm}   






\begin{thebibliography}{99}

\bibitem{c1}van der Zee, S. C., Fischer, P. H., and Hoek, G. (2016). Air pollution in perspective: Health risks of air pollution expressed in equivalent numbers of passively smoked cigarettes. Environmental Research, 148, 475–483.
\bibitem{c2}Caiazzo, F., Ashok, A., Waitz, I. A., Yim, S. H. L., and Barrett, S. R. H. (2013). Air pollution and early deaths in the United States. Part I: Quantifying the impact of major sectors in 2005. Atmospheric Environment, 79, 198–208. 
\bibitem{c3}Gurjar, B. R., Jain, A., Sharma, A., Agarwal, A., Gupta, P., Nagpure, A. S., and Lelieveld, J. (2010). Human health risks in megacities due to air pollution. Atmospheric Environment, 44(36), 4606–4613. https://doi.org/10.1016/J.ATMOSENV.2010.08.011
\bibitem{c4}Zevin S, Saunders S, Gourlay SG, Jacob P, Benowitz NL. Cardiovascular effects of carbon monoxide and cigarette smoking. J Am Coll Cardiol. 2001;38(6):1633-1638. doi:10.1016/S0735-1097(01)01616-3
\bibitem{c5}Wasserman, K., Whipp, B. J., Koyl, S. N., and Beaver, W. L. (1973). Anaerobic threshold and respiratory gas exchange during exercise. Journal of Applied Physiology, 35(2), 236–243. https://doi.org/10.1152/jappl.1973.35.2.236

\bibitem{c6}"Strava | Run and Cycling Tracking on the Social Network for Athletes", Strava.com, 2018. [Online]. Available: https://www.strava.com/. [Accessed: 16- Aug- 2018]

\bibitem{c7}Cottin, F., Papelier, Y., and Escourrou, P. (1999). Effects of Exercise Load and Breathing Frequency on Heart Rate and Blood Pressure Variability During Dynamic Exercise. International Journal of Sports Medicine, 20(04), 232–238. https://doi.org/10.1055/s-2007-971123

\bibitem{c8}Bascom, R., Bromberg, P. A., Costa, D. L., Devlin, R., Dockery, D. W., Frampton, M. W., … Utell, M. (1996). Health effects of outdoor air pollution. Part 2. Committee of the Environmental and Occupational Health Assembly of the American Thoracic Society. American Journal of Respiratory and Critical Care Medicine, 153(2), 477–498. https://doi.org/10.1164/ajrccm.153.2.8564086
\bibitem{c9}"OpenAQ", OpenAQ, 2018. [Online]. Available: https://openaq.org/. [Accessed: 16- Aug- 2018]
\bibitem{c10}"Ambient air pollution: Health impacts", World Health Organization, 2018. [Online]. Available: http://www.who.int/airpollution/ambient/health-impacts/en/. [Accessed: 21- Aug- 2018]
\bibitem{c11}Yuan, George, Nicole A. Drost, and R. Andrew McIvor. "Respiratory rate and breathing pattern." McMaster University Medical Journal 10, no. 1 (2013): 23-25.
\bibitem{c12}Natalini G, Marchesini M, Tessadrelli A, Rosano A, Candiani A, Bernardini A. Effect of tidal volume and respiratory rate on the power of breathing calculation. Acta Anaesthesiol Scand. 2005;49(5):643-648. doi:10.1111/j.1399-6576.2005.00664.x
\bibitem{c13}Caiazzo F, Ashok A, Waitz IA, Yim SHL, Barrett SRH. Air pollution and early deaths in the United States. Part I: Quantifying the impact of major sectors in 2005. Atmos Environ. 2013;79:198-208. doi:10.1016/j.atmosenv.2013.05.081
\bibitem{c14} van der Zee SC, Fischer PH, Hoek G. Air pollution in perspective: Health risks of air pollution expressed in equivalent numbers of passively smoked cigarettes. Environ Res. 2016;148:475-483. doi:10.1016/J.ENVRES.2016.04.001
\bibitem{c15}Brower RG, Matthay MA, Morris A, Schoenfeld D, Thompson BT, Wheeler A. Ventilation with lower tidal volumes as compared with traditional tidal volumes for acute lung injury and the acute respiratory distress syndrome.
Acute Respiratory Distress Syndrome Network, N Engl J Med. 2000 May 4;342(18):1301-8.
\bibitem{c16}Watson AWS. The Relationship Between Tidal Volume and Respiratory Frequency During Muscular Exercise. Br J Sports Med. 2008;8(2-3):87-90. doi:10.1136/bjsm.8.2-3.87
\bibitem{c17}Nitish N, Pandey V, Bhimaraju H, et al. Cross-Modal Health State Estimation. ACM Multimed. August 2018:1993-2002. doi:10.1145/3240508.3241913
\bibitem{c18}Nag N, Jain R. A Navigational Approach to Health. IEEE Comput. December 2018. doi:10.1109/MC.2018.2883280
\bibitem{c19}Jiang C, Wang X, Li X, et al. Dynamic Human Environmental Exposome Revealed by Longitudinal Personal Monitoring. Cell. 2018;175(1):277-291.e31. doi:10.1016/j.cell.2018.08.060
\end{thebibliography}
\end{document}